\documentclass[11pt,twocolumn]{article}
\usepackage{graphicx}
\usepackage{amsmath}
\usepackage{bm}

\begin{document}

\title{\textsf{\textbf{Paraxial meridional ray tracing equations from the unified reflection-refraction law via geometric algebra}}}
\author{Quirino M. Sugon Jr.* and Daniel J. McNamara
\smallskip\\
\small{Ateneo de Manila University, Department of Physics, Loyola Heights, Quezon City, Philippines 1108}\\
\small{*Also at Manila Observatory, Upper Atmosphere Division, Ateneo de Manila University Campus}\\
\small{e-mail: \texttt{qsugon$@$observatory.ph}}}
\date{\small{\today}}
\maketitle

\section*{}
\small{\textbf{Abstract.}} 
We derive the paraxial meridional ray tracing equations from the unified reflection-refraction law using geometric algebra.  This unified law states that the normal vector to the interface is a rotation of the incident ray or of the refracted ray or of the reflected ray by an angle equal to the angle of incidence or of refraction.  We obtain the finite meridional ray tracing equations by simply equating the arguments of the exponential rotation operators. We then derive the paraxial limits of these equations with the help of sign function identities.  We show that by embedding the sign functions in the ray tracing equations, we explicitly declare our chosen sign conventions in symbols and not in prose.

\section{Introduction}

In paraxial optics, it is customary to declare beforehand the adopted set of sign conventions, as done for example in Nussbaum and Philips\cite{NussbaumPhilips_1976_ContemporaryOpticsforScientistsandEngineers_p11-12}.  But there are as many sign conventions as there matrix optics authors.  And debate ensues\cite{Sukheeja_1967_ajp35i7pp611-613}-\cite{Nussbaum_2000_ajp68i4p305}.  To get a taste of this controversy, let us quote Welford's  1974 critique of Conrady's convention\cite{Welford_1974_AbberationsoftheSymmetricalOpticalSystem_pvi}\cite{Conrady_1985_AppliedOpticsandOpticalDesign}:
\begin{quote}
\footnotesize
I have avoided the issue [of sign conventions] by simply using the universally accepted conventions of coordinate geometry, together with vectors and direction cosines, as a consistent system which agrees with what is done in other branches of physics.  This conflicts with what has been taught at Imperial College on one small point: the sign of the paraxial convergence angle, $u$, has always been taken according to Conrady's convention but after due discussing with my colleagues I decided to reverse it so as to agree with the convention for direction cosines; the inconsistency could, we felt, no longer be justified.
\end{quote}

To resolve this sign problem, we propose the use of sign functions that take values of $\pm 1$, such as the three axial direction functions for the vector $\mathbf v$:
\begin{eqnarray}
\label{eq:c_vx intro}
c_{vx}&=&\frac{\mathbf v\cdot\mathbf e_1}{|\mathbf v\cdot\mathbf e_1|},\\
\label{eq:c_vy intro}
c_{vy}&=&\frac{\mathbf v\cdot\mathbf e_2}{|\mathbf v\cdot\mathbf e_2|},\\
\label{eq:c_vz intro}
c_{vz}&=&\frac{\mathbf v\cdot\mathbf e_3}{|\mathbf v\cdot\mathbf e_3|}.
\end{eqnarray}
These functions correspond to the signs of the direction cosines of a vector $\mathbf v$.  So by using these sign functions, we explicitly adopt Welford's Cartesian sign conventions.

Another example of a sign function is the concavity function\cite{SugonMcNamara_2004_ajp72i1pp92-97_p95}:
\begin{equation}
\label{eq:c_sigma eta intro}
c_{\sigma\eta}=\frac{\bm\sigma\cdot\bm\eta}{|\bm\sigma\cdot\bm\eta|},
\end{equation}
where $\bm\sigma$ is the incident ray and $\bm\eta$ is the normal vector to the interface.  If the interface is concave, then $c_{\sigma\eta}=1$; if convex, $c_{\sigma\eta}=-1$.  We used the concavity function before when we wrote down the unified reflection-refraction law in exponential form\cite{SugonMcNamara_2006_AIEP139pp179-224_p192-195}: 
\begin{equation}
\label{eq:unified reflection refraction law intro}
c_{\sigma\eta}\bm\eta=\bm\sigma e^{ic_{\sigma\eta}\beta\mathbf e_{\sigma\times\eta}}=\bm\sigma' e^{ic_{\sigma\eta}\beta'\mathbf e_{\sigma\times\eta}}=-\bm\sigma''e^{-ic_{\sigma\eta}\beta''\mathbf e_{\sigma\times\eta}},
\end{equation}
where $\bm\sigma'$ is the refracted ray and $\bm\sigma''$ is the reflected ray.  From this unified law we derived the ray tracing equations for finite and paraxial skew rays in spherical coordinates and for finite meridional rays in polar coordinates.  The paraxial meridional rays we deemed then to require a separate treatment; we set it aside for a future work.

In this paper, we shall continue our work.  We shall start with a short review of geometric algebra and then proceed to geometric optics.  We shall summarize the equations for finite skew and finite meridional ray tracing, and then use these to derive those for paraxial meridional rays in polar coordinates.  We shall see how the use of sign functions makes the discussion of sign conventions unncessary.

\section{Geometric Algebra}

In Clifford (geometric) algebra $\mathcal Cl_{3,0}$ the product of two vectors ${\bf a}$ and ${\bf b}$ is given by the Pauli identity\cite{Hestenes_2003_ajp71i2pp104-121_p110}:
\begin{equation}
\label{eq:Pauli identity}
{\bf a}{\bf b} = {\bf a}\cdot{\bf b} + i({\bf a}\times{\bf b}),
\end{equation}
where $i$ is the unit imaginary scalar.  Note that the geometric product $\mathbf a\mathbf b$ is an associative product, unlike the dot product $\mathbf a\cdot\mathbf b$ and the cross product $\mathbf a\times\mathbf b$.

The exponential function in geometric algebra is also well-defined:
\begin{equation}
\label{eq:Euler theorem}
e^{i\bm\theta}=\cos|{\bm\theta}|+i\frac{\bm\theta}{|{\bm\theta}|}\sin|{\bm\theta}|,
\end{equation}
which is the generalization of Euler's theorem in complex analysis.  If $\mathbf a$ is a vector perpendicular to $\bm\theta$, then we can show that  
\begin{eqnarray}
\label{eq:a exp i theta is a cos theta + i a theta sin theta}
\mathbf ae^{i\bm\theta}=\mathbf a\cos|{\bm\theta}|-\mathbf a\times\frac{\bm\theta}{|\bm\theta|}\sin|\bm\theta|.
\end{eqnarray} 
Equation~(\ref{eq:a exp i theta is a cos theta + i a theta sin theta}) states that $\mathbf ae^{i\bm\theta}$ is the vector $\mathbf a$ rotated counterclockwise about the vector $\bm\theta$ by an angle $|\bm\theta|$.\cite{SugonMcNamara_2004_ajp72i1pp92-97_p93}

\section{Geometric Optics}

\subsection{Finite Skew Rays}

Finite skew rays\cite{SugonMcNamara_2006_AIEP139pp179-224_p188-196} are the most general type of rays.  The ray tracing equations for these rays are expressed in vector form .

From its initial position $\mathbf r_0$, a light particle travels by a distance $s$ in the direction of the unit vector $\bm\sigma$.  The final position $\mathbf r$ of the light particle is
\begin{equation}
\label{eq:propagation}
\mathbf r=\mathbf r_0+s\bm\sigma.
\end{equation}
If at the position $\mathbf r$, the outward normal unit vector to the interface is $\bm\eta$ and the interface is spherical of radius $R$ centered at $C$, then
\begin{equation}
\label{eq:propagation intersection}
\mathbf r=\mathbf r_0+s\bm\sigma=\mathbf C+R\bm\eta.
\end{equation} 

After the light particle strikes the interface, the particle is either be refracted or reflected.  The directions $\bm\sigma^\prime$ and $\bm\sigma^{\prime\prime}$ of the refracted and reflected vectors may be expressed in terms of the directions $\bm\sigma$ of the incident ray and (outward) normal vector $\bm\eta$ to the interface:
\begin{eqnarray}
\label{eq:refraction law asymmetric}
\bm\sigma^\prime&=&\bm\sigma e^{i(\beta-\beta^\prime)\mathbf e_{\sigma\!\times\!\eta}},\\
\label{eq:reflection law asymmetric}
\bm\sigma^{\prime\prime}&=&\bm\sigma e^{2ic_{\sigma\eta}\beta\mathbf e_{\sigma\times\eta}},
\end{eqnarray}
where $\beta$ and $\beta^\prime$ are the angles of incidence and refraction,
\begin{eqnarray}
\label{eq:beta is asin magnitude sigma x eta}
\beta&=&\sin^{-1}|\bm\sigma\times\bm\eta|,\\
\beta^\prime&=&\sin^{-1}|\bm\sigma\times\bm\eta|,
\end{eqnarray}
and $\mathbf e_{\sigma\times\eta}$ is the rotational axis direction,  
\begin{equation}
\label{eq:e_sigmaxeta}
\mathbf e_{\sigma\times\eta} =\frac{\bm\sigma\times\bm\eta}{|\bm\sigma\times\bm\eta|}.
\end{equation}

We may also combine the laws of refraction and reflection in Eqs.~(\ref{eq:refraction law asymmetric}) and (\ref{eq:reflection law asymmetric}) into one, as given in Eq.~(\ref{eq:unified reflection refraction law intro}).  This unified law expresses the normal vector in terms of the rotations of the incident, refracted, and reflected rays about the vector $\mathbf e_{\sigma\times\eta}$ by angles $\beta$, $\beta^\prime$, and $\beta^{\prime\prime}=\beta$, respectively.  The rotations are counterclockwise or clockwise depending on the sign value of the concavity function $c_{\sigma\eta}$.  (Figure 1)

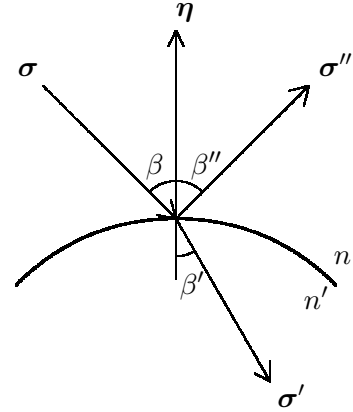
\begin{figure}[h]
\begin{center}
\setlength{\unitlength}{1 mm}
\begin{picture}(60,65)(-30,-30)
\thicklines
\qbezier(21.213,-8.787)(12.426,0.000)(0.000,0.000)
\qbezier(0.000,0.000)(-12.426,0.000)(-21.213,-8.787)
\put(17,-12){$n'$}
\put(21,-6){$n$}
\thinlines
\qbezier(-17.678,17.678)(0,0)(0,0)
\qbezier(-2.5,0.5)(0,0)(0,0)
\qbezier(-0.7,2.5)(0,0)(0,0)
\put(-21,19){$\bm\sigma$}
\qbezier(0,-8)(0,25)(0,25)
\qbezier(-1,23)(0,25)(0,25)
\qbezier(1,23)(0,25)(0,25)
\put(0,27){$\bm\eta$}
\qbezier(0,0)(17.678,17.678)(17.678,17.678)
\qbezier(17,15)(17.678,17.678)(17.678,17.678)
\qbezier(15,17)(17.678,17.678)(17.678,17.678)
\put(19,19){$\bm\sigma''$}
\qbezier(0,0)(12.500,-21.651)(12.500,-21.651)
\qbezier(12.5,-19)(12.500,-21.651)(12.500,-21.651)
\qbezier(10.5,-20.6)(12.500,-21.651)(12.500,-21.651)
\put(13.5,-25){$\bm\sigma'$}
\qbezier(3.536,3.536)(2.071,5.000)(0.000,5.000)
\qbezier(0.000,5.000)(-2.071,5.000)(-3.536,3.536)
\put(-4,6){$\beta$}
\put(2,6){$\beta''$}
\qbezier(2.500,-4.330)(1.340,-5.000)(0.000,-5.000)
\put(0.5,-10){$\beta'$}
\end{picture}
\end{center}
\begin{quote}
\vspace{-0.5cm}
\caption{\footnotesize The incident ray $\bm\sigma$, refracted ray $\bm\sigma'$, and reflected ray $\bm\sigma''$.  The rays make an angle of $\beta$, $\beta'$, and $\beta''$ with respect to the normal vector $\bm\eta$, respectively.  The interface is convex, $c_{\sigma\eta}=-1$.}
\label{fig:incident refracted reflected rays}
\vspace{-1cm}
\end{quote}
\end{figure}

\subsection{Finite Meridional Rays}

Finite meridional rays \cite{SugonMcNamara_2006_AIEP139pp179-224_p215-221} are rays that lie on the same plane.  Here, we choose this plane to be the $zx-$plane, with the $z-$axis along $\mathbf e_3$ as the optical axis.

Let us define the incident, refracted, reflected, and normal vectors as vectors in $zx$-plane:
\begin{eqnarray}
\label{eq:incident ray polar}
\bm\sigma&=&\mathbf e_3e^{i\mathbf e_2\theta_\sigma}=\mathbf e_3\cos\theta_\sigma+\mathbf e_1\sin\theta_\sigma,\\
\label{eq:refracted ray polar}
\bm\sigma^\prime&=&\mathbf e_3e^{i\mathbf e_2\theta_{\sigma^\prime}}=\mathbf e_3\cos\theta_{\sigma^\prime}+\mathbf e_1\sin\theta_{\sigma^\prime},\\
\label{eq:reflected ray polar}
\bm\sigma^{\prime\prime}&=&\mathbf e_3e^{i\mathbf e_2\theta_{\sigma^{\prime\prime}}}=\mathbf e_3\cos\theta_{\sigma^{\prime\prime}}+\mathbf e_1\sin\theta_{\sigma^{\prime\prime}},\\
\label{eq:normal vector polar}
\bm\eta&=&\mathbf e_3e^{i\mathbf e_2\theta_\eta}=\mathbf e_3\cos\theta_\eta+\mathbf e_1\sin\theta_\eta,
\end{eqnarray}
where $\theta$ is a counterclockwise rotation angle measured from $\mathbf e_3$.

If we also define the position vectors $\mathbf r$ and $\mathbf r_0$ as vectors in the $zx-$plane,
\begin{eqnarray}
\label{eq:position r polar}
\mathbf r&=&z\mathbf e_3+x\mathbf e_1,\\
\label{eq:position r0 polar}
\mathbf r_0&=&z_0\mathbf e_3+x_0\mathbf e_1,
\end{eqnarray}
and the center 
\begin{equation}
\label{eq:center C}
\mathbf C=z_C\mathbf e_3
\end{equation}
of the interface to be along the optical axis $\mathbf e_3$, then Eq.~(\ref{eq:propagation intersection}) separates into
\begin{eqnarray}
\label{eq:propagation intersection z part}
z&=&z_0+s\cos\theta_\sigma=C+R\cos\theta_\eta,\\
\label{eq:propagation intersection x part}
x&=&x_0+s\sin\theta_\sigma=R\sin\theta_\eta.
\end{eqnarray}
Equations~(\ref{eq:propagation intersection z part}) and (\ref{eq:propagation intersection x part}) are the ray propagation equations for finite meridional rays. (Figure 2)

On the other hand, substituting the definitions in Eqs.~(\ref{eq:incident ray polar}) to (\ref{eq:normal vector polar}) back to the unified reflection-refraction law in Eq.~(\ref{eq:unified reflection refraction law intro}) and employing the identities
\begin{eqnarray}
\label{eq:e_sigma times eta is c_sigma times eta y e_2}
\mathbf e_{\sigma\times\eta}&=&c_{(\sigma\times\eta)y}\mathbf e_2,\\
\label{eq:concavity function as exponential}
c_{\sigma\eta}&=&e^{i\mathbf e_2(c_{\sigma\eta}-1)\pi/2},\\
\label{eq:-1 as exponential}
-1&=&e^{i\mathbf e_2\pi}
\end{eqnarray}
we arrive at
\begin{eqnarray}
\label{eq:unified refraction-reflection law arguments}
\frac{\pi}{2}(c_{\sigma\eta}-1)+\theta_\eta&=&\theta_\sigma+c_{\sigma\eta}c_{(\sigma\times\eta)y}\beta,\nonumber\\
&=&\theta_{\sigma^\prime}+c_{\sigma\eta}c_{(\sigma\times\eta)y}\beta^\prime,\nonumber\\
&=&\pi+\theta_{\sigma^{\prime\prime}}-c_{\sigma\eta}c_{(\sigma\times\eta)y}\beta.
\end{eqnarray}
Equation~(\ref{eq:unified refraction-reflection law arguments}) is the unified reflection-refraction law for finite meridional rays.  We can show that this equation contains a restatement of the Bessel-Conrady refraction invariant, by replacing $\theta_\sigma$ by $U$ and $c_{\sigma\eta}c_{(\sigma\times\eta)y}\beta$ by I. \cite{Conrady_1985_AppliedOpticsandOpticalDesign_p7}\cite{SugonMcNamara_2006_AIEP139pp179-224_p219}

\begin{figure}[hb]
\begin{center}
\setlength{\unitlength}{1 mm}
\begin{picture}(80,50)(10,-10)
\put(10,0){\line(1,0){80}}
\put(20,-2.5){\line(0,1){17.5}}
\put(19,-6){$z_0$}
\put(22,5){$x_0$}
\put(50,-2.5){\line(0,1){32.5}}
\put(49,-6){$z$}
\put(52,5){$x$}
\qbezier(20,15)(50,30)(50,30)
\qbezier(48,30)(50,30)(50,30)
\qbezier(49,28)(50,30)(50,30)
\put(35,25){$s$}
\put(20,15){\line(1,0){8}}
\qbezier(25.000,15.000)(25.000,16.340)(24.330,17.500)
\put(29,16){$\theta_\sigma$}
\thicklines
\qbezier(52.729,32.501)(35.810,18.304)(37.735,-3.698)
\thinlines
\qbezier(80,0)(50,30)(50,30)
\put(67,15){$R$}
\qbezier(83.000,0.000)(83.000,1.820)(81.385,2.661)
\qbezier(81.385,2.661)(79.434,3.677)(77.879,2.121)
\put(81,4){$\theta_\eta$}
\put(80,0){\line(0,-1){2.5}}
\put(79,-6){$z_C$}
\end{picture}
\end{center}
\begin{quote}
\vspace{-0.5cm}
\caption{\footnotesize A ray travels a distance $s$ until it intersects an interfaces of radius $R$.}
\label{fig:propagating ray intersects an interface}
\vspace{-1cm}
\end{quote}
\end{figure}
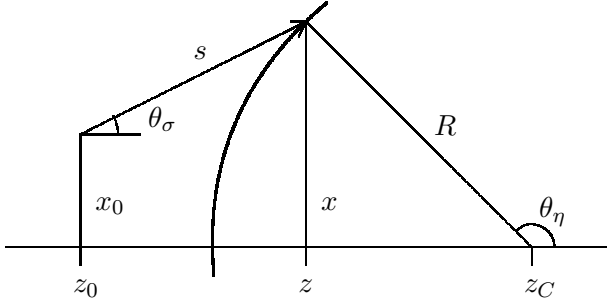

\subsection{Paraxial Meridional Rays}

Paraxial meridional rays are rays that lie on the same plane ($zx-$plane) and make a small angle with respect to the optical axis $\mathbf e_3$.  Mathematically, we say that if $\mathbf v$ is a paraxial meridional ray, then the polar angle $\theta_{v}$ of $\mathbf v$ may be expressed in terms of the small positive angle $\theta_{vz}$ that $\mathbf v$ makes with the optical axis\cite{SugonMcNamara_2006_AIEP139pp179-224_p204}:
\begin{equation}
\label{eq:theta_v}
\theta_{v}=\frac{\pi}{2}(1-c_{vz})+c_{vz}c_{vx}\theta_{vz},
\end{equation}
where $\theta_v$ is the polar angle of $\mathbf v$.  We can easily verify that
\begin{eqnarray}
\label{eq:cos theta_v}
\cos\theta_v&=&c_{vz},\\
\label{eq:sin theta_v}
\sin\theta_v&=&c_{vx}\theta_{vz}.
\end{eqnarray}
Note that $c_{vx}$ and $c_{vz}$ are the axial direction functions defined in Eqs.~(\ref{eq:c_vx intro}) and (\ref{eq:c_vz intro}).
\subsubsection{Propagation}

Using the approximations in Eqs.~(\ref{eq:cos theta_v}) and (\ref{eq:sin theta_v}), Eqs.~(\ref{eq:propagation intersection z part}) and (\ref{eq:propagation intersection x part}) simplify to
\begin{eqnarray}
\label{eq:propagation intersection z part paraxial}
z&=&z_0+c_{\sigma z}s=C+c_{\eta z}R,\\
\label{eq:propagation intersection x part paraxial}
x&=&x_0+c_{\sigma x}s\,\theta_{\sigma z}=c_{\eta x}R\theta_{\eta z}.
\end{eqnarray}
Equations~(\ref{eq:propagation intersection z part paraxial}) and (\ref{eq:propagation intersection x part paraxial}) are the position-height relations for paraxial meridional ray propagation.

For the polar angle $\theta_{\sigma}$, we know that it is conserved during propagation or translation:
\begin{equation}
\label{eq:theta sigma is theta sigma 0}
\theta_\sigma=\theta_{\sigma 0}.
\end{equation}
Using Eq.~(\ref{eq:theta_v}), Eq.~(\ref{eq:theta sigma is theta sigma 0}) may be expanded as
\begin{equation}
\label{eq:theta sigma is theta sigma 0 expand}
\frac{\pi}{2}(1-c_{\sigma z})+c_{\sigma z}c_{\sigma x}\theta_{\sigma z}=\frac{\pi}{2}(1-c_{\sigma 0z})+c_{\sigma 0z}c_{\sigma 0x}\theta_{\sigma 0z},
\end{equation}
Because the light particle moves in the same direction of the z-axis during propagation, then 
\begin{equation}
\label{eq:c_sigma z is c_sigma 0z}
c_{\sigma z}=c_{\sigma 0z}
\end{equation}
so that Eq.~(\ref{eq:theta sigma is theta sigma 0 expand}) reduces to
\begin{equation}
\label{eq:theta sigma is theta sigma 0 expand reduced}
c_{\sigma x}\theta_{\sigma z}=c_{\sigma 0x}\theta_{\sigma 0z}.
\end{equation}
Thus, the inclination angle of a ray from the optical axis remains invariant under ray propagation.

Substituting Eq.~(\ref{eq:theta sigma is theta sigma 0 expand reduced}) back to the propagation equation in Eq.~(\ref{eq:propagation intersection x part paraxial}), we obtain
\begin{equation}
\label{eq:x is x_0 + s c_sigma 0x theta_sigma 0z}
x=x_0+sc_{\sigma0x}\theta_{\sigma0 z}.
\end{equation}
Equation~(\ref{eq:x is x_0 + s c_sigma 0x theta_sigma 0z}) expresses the height $x$ of a ray from the optical axis as a function of the ray's initial height $x_0$, distance travelled $s$, and direction angle $\theta_{\sigma 0z}$.  Notice that the magnitude $\theta_{\sigma0 z}$ of the paraxial angle is divorced from its $x$-direction sign function $c_{\sigma 0x}$.

\subsubsection{Refraction}

The refraction law may be extracted from the unified law in Eq.~(\ref{eq:unified refraction-reflection law arguments}):
\begin{equation}
\label{eq:refraction law finite meridional}
\theta_{\sigma^\prime}=\theta_\sigma+c_{\sigma\eta}c_{(\sigma\times\eta)y}(\beta-\beta^\prime).
\end{equation}
We know that the angles of incidence and refraction are related by Descartes-Snell's law:
\begin{equation}
\label{eq:Descartes-Snell law}
n^\prime\sin\beta^\prime=n\sin\beta,
\end{equation}
where $n$ and $n^\prime$ are the refractive indices of the medium containing the incident and refracted rays, respectively.  In the paraxial limit, Eq.~(\ref{eq:Descartes-Snell law}) becomes
\begin{equation}
\label{eq:Descartes-Snell law paraxial}
n^\prime\beta^\prime=n\beta.
\end{equation}
Hence,
\begin{equation}
\label{eq:refraction law paraxial}
\theta_{\sigma^\prime}=\theta_\sigma+c_{\sigma\eta}c_{(\sigma\times\eta)y}(1-\mu)\beta.
\end{equation}
where
\begin{equation}
\label{eq:mu}
\mu=\frac{n}{n^\prime}.
\end{equation}
Equation~(\ref{eq:refraction law paraxial}) is the paraxial refraction law in terms of the polar angle $\theta_\sigma$ of the incident ray and the angle of incidence $\beta$.
            		 
Using the unified refraction-reflection law in Eq.~(\ref{eq:unified refraction-reflection law arguments}), we solve for the angle of incidence $\beta$ in terms of the polar angle $\theta_\eta$ of the normal vector:
\begin{equation}
\label{eq:incident angle beta is theta_eta - theta_sigma}
c_{\sigma\eta}c_{(\sigma\times\eta)y}\beta=\frac{\pi}{2}(c_{\sigma\eta}-1)+\theta_\eta-\theta_\sigma,
\end{equation}
Substituting this back to Eq.~(\ref{eq:refraction law paraxial}) and rearranging the terms, we arrive at	
\begin{equation}
\label{eq:refraction law paraxial beta replaced}
\theta_{\sigma^\prime}=\mu\theta_\sigma+(1-\mu)(\frac{\pi}{2}(c_{\sigma\eta}-1)+\theta_\eta).
\end{equation}
In paraxial approximation, this is
\begin{eqnarray}
\label{eq:refraction law paraxial beta replaced paraxial expand}
\frac{\pi}{2}(1-c_{\sigma^\prime z})&+&c_{\sigma^\prime z}c_{\sigma^\prime x}\theta_{\sigma^\prime z}\nonumber\\
&=&\mu(1-c_{\sigma z})\frac{\pi}{2}+\mu c_{\sigma z}c_{\sigma x}\theta_{\sigma z}\nonumber\\
& &+\,(1-\mu)(c_{\sigma\eta}-c_{\eta z})\frac{\pi}{2}\nonumber\\
& &+\,(1-\mu)c_{\eta z}c_{\eta x}\theta_{\eta z}.
\end{eqnarray}

Now, we can verify that the following sign identities hold:
\begin{eqnarray}
\label{eq:c_sigma prime z is c_sigma z}
c_{\sigma^\prime z}&=&c_{\sigma z},\\
\label{eq:c_sigma eta is c_sigma c_eta}
c_{\sigma\eta}&=&c_{\sigma z} c_{\eta z}.
\end{eqnarray}			
The first equation states that the relative directions of the incident and refracted rays with respect to the optical axis are the same; the second states that the concavity function is equal to the product of the relative directions of the incident ray and normal vector with respect to the optical axis.

Employing the sign identities in Eqs.~(\ref{eq:c_sigma prime z is c_sigma z}) and (\ref{eq:c_sigma eta is c_sigma c_eta}), Eq.~(\ref{eq:refraction law paraxial beta replaced paraxial expand}) reduces to
\begin{eqnarray}
\label{eq:refraction law paraxial beta replaced paraxial reduced}
c_{\sigma^\prime x}\theta_{\sigma^\prime z}&=&m'+\mu c_{\sigma x}\theta_{\sigma z}+(1-\mu)c_{\sigma\eta}c_{\eta x}\theta_{\eta z},
\end{eqnarray}
where
\begin{equation}
\label{eq:m' function}
m'=(1-\mu)(c_{\eta z}-c_{\sigma z}c_{\eta z}-c_{\sigma z}+1)\frac{\pi}{2}.
\end{equation}

Let us analyze the angular function $m'$ in Eq.~(\ref{eq:m' function}) by considering two cases:
\begin{eqnarray}
\label{eq:m' function case +1}
m'&=&0;\quad\quad\quad\quad\quad\quad\quad\quad c_{\sigma z}=+1,\\
\label{eq:m function case -1}
m'&=&\pi(1-\mu)(1+c_{\eta z});\quad c_{\sigma z}=-1.
\end{eqnarray}
Because the direction $c_{\eta z}$ of the normal vector with respect to the $\mathbf e_3$ is arbitrary, then $m'\neq 0$ in general for backward propagating rays $(c_{\sigma z}=-1)$.  This is an inconvenient case.  So we shall leave this for a future work and impose that
\begin{equation}
\label{eq:c_sigma z is +1}
c_{\sigma z}=c_{\sigma^\prime z}=+1
\end{equation}
for all our equations.  Thus, Eq.~(\ref{eq:refraction law paraxial beta replaced paraxial reduced}) reduces further to
\begin{equation}
\label{eq:refraction law paraxial forward}
c_{\sigma^\prime x}\theta_{\sigma^\prime z}=\mu c_{\sigma x}\theta_{\sigma z}+(1-\mu)c_{\eta z}c_{\eta x}\theta_{\eta z}.
\end{equation}
Equation~(\ref{eq:refraction law paraxial forward}) is the 
refraction law for forward propagating paraxial rays.

Using the result in Eq.~(\ref{eq:propagation intersection x part paraxial}),
\begin{equation}
\label{eq:theta_eta z as x over R}
c_{\eta x}\theta_{\eta z}=\frac{x}{R},
\end{equation}
Eq.~(\ref{eq:refraction law paraxial forward}) becomes
\begin{equation}
\label{eq:refraction law paraxial forward x R}
c_{\sigma^\prime x}\theta_{\sigma^\prime z}=\mu c_{\sigma x}\theta_{\sigma z}+(1-\mu)c_{\eta z}\frac{x}{R}.
\end{equation}
Multiplying this by $n^\prime$ yields
\begin{equation}
\label{eq:refraction law paraxial forward x P}
n^\prime c_{\sigma^\prime x}\theta_{\sigma^\prime z}=nc_{\sigma x}\theta_{\sigma z}-Px,
\end{equation}
where
\begin{equation}
\label{eq:power P}
P=c_{\eta z}\frac{n-n^\prime}{R}
\end{equation}
is the power of the interface.  Notice that the radius $R$ of the interface is always positive; the $z-$axis direction function $c_{\eta z}$ of the normal vector $\bm\eta$ takes care of the sign traditionally possessed by $R$.
	
\subsubsection{Reflection}

From the unified refraction-reflection law in Eq.~(\ref{eq:unified refraction-reflection law arguments}), we get
\begin{equation}
\label{eq:reflection law finite meridional}
\theta_{\sigma^{\prime\prime}}=-\pi+\theta_\sigma+2c_{\sigma\eta}c_{(\sigma\times\eta)y}\beta.
\end{equation}
Using the result in Eq.~(\ref{eq:incident angle beta is theta_eta - theta_sigma}), Eq.~(\ref{eq:reflection law finite meridional}) becomes
\begin{equation}
\label{eq:reflection law finite meridional no beta}
\theta_{\sigma^{\prime\prime}}=(c_{\sigma\eta}-2)\pi-\theta_\sigma+2\theta_\eta.
\end{equation}

In the paraxial limit, Eq.~(\ref{eq:reflection law finite meridional no beta}) reduces to
\begin{eqnarray}
\label{eq:reflection law paraxial expand}
\frac{\pi}{2}(1-c_{\sigma^{\prime\prime}z})&+& c_{\sigma^{\prime\prime}z}c_{\sigma^{\prime\prime}x}\theta_{\sigma^{\prime\prime}z}\\
&=&\pi(c_{\sigma\eta}-c_{\eta z}-1)+\frac{\pi}{2}(c_{\sigma z}-1)\nonumber\\
& &-\ c_{\sigma z}c_{\sigma x}\theta_{\sigma z}+2c_{\eta z}c_{\eta x}\theta_{\eta z}.
\end{eqnarray}
Because the $z-$direction of the reflected ray is opposite to that of the incident ray, then
\begin{equation}
\label{eq:c_sigma prime prime is -c_sigma}
c_{\sigma^{\prime\prime}z}=-c_{\sigma z},
\end{equation}
so that Eq.~(\ref{eq:reflection law paraxial expand}) becomes
\begin{eqnarray}
\label{eq:reflection law paraxial reduced}
c_{\sigma^{\prime\prime}x}\theta_{\sigma^{\prime\prime}z}&=&\pi(-c_{\eta z}+c_{\sigma z}c_{\eta z}+2c_{\sigma z})\nonumber\\
& &+\ c_{\sigma x}\theta_{\sigma z}+2c_{\sigma\eta}c_{\eta x}\theta_{\eta z},
\end{eqnarray}
after multiplying by $c_{\sigma z}$ and using the sign identity in Eq.~(\ref{eq:c_sigma eta is c_sigma c_eta}).

Let us analyze the angular function
\begin{equation}
\label{eq:m prime prime function}
m^{\prime\prime}=\pi(-c_{\eta z}+c_{\sigma z}c_{\eta_z}+2c_{\sigma z})
\end{equation}
by considering two cases:
\begin{eqnarray}
\label{eq:m prime prime case +1}
m^{\prime\prime}&=&2\pi;\ \ \qquad\qquad\qquad\qquad\qquad c_{\sigma z}=+1,\\
\label{eq:m prime prime case -1}
m^{\prime\prime}&=&-2\pi(c_{\eta z}+1)=\{0,-4\pi\};\quad c_{\sigma z}=-1,
\end{eqnarray}
since $c_{\eta z}=\pm 1$.  Because $4\pi\equiv 2\pi\equiv 0$, then we may simply set $m^{\prime\prime}=0$, so that Eq.~(\ref{eq:reflection law paraxial reduced}) simplifies further to
\begin{equation}
\label{eq:reflection law paraxial}
c_{\sigma^{\prime\prime}x}\theta_{\sigma^{\prime\prime}z}=c_{\sigma x}\theta_{\sigma z}+2c_{\sigma\eta}c_{\eta x}\theta_{\eta z}.
\end{equation}
Employing the identity in Eq.~(\ref{eq:theta_eta z as x over R}), Eq.~(\ref{eq:reflection law paraxial}) becomes
\begin{equation}
\label{eq:reflection law paraxial x R}
c_{\sigma^{\prime\prime}x}\theta_{\sigma^{\prime\prime}z}=c_{\sigma x}\theta_{\sigma z}-2c_{\sigma\eta}\frac{x}{R}.
\end{equation}

If we impose that the incident ray moves along $\mathbf e_3$, then $c_{\sigma z}=+1$, so that Eq.~(\ref{eq:reflection law paraxial x R}) simplifies to
\begin{equation}
\label{eq:reflection law paraxial x R +1}
c_{\sigma^{\prime\prime}x}\theta_{\sigma^{\prime\prime}z}=c_{\sigma x}\theta_{\sigma z}-2c_{\eta z}\frac{x}{R}.
\end{equation}
Multiplying Eq.~(\ref{eq:reflection law paraxial x R +1}) by $n^{\prime\prime}=n$,
\begin{equation}
\label{eq:reflection law paraxial x R +1 n}
n^{\prime\prime}c_{\sigma^{\prime\prime}x}\theta_{\sigma^{\prime\prime}z}=nc_{\sigma x}\theta_{\sigma z}-2nc_{\eta z}\frac{x}{R},
\end{equation}
and comparing the result with the refraction law in Eq.~(\ref{eq:refraction law paraxial forward x P}), we see that we may express the mirror power $P^{\prime\prime}$ as
\begin{equation}
\label{eq:power P prime prime}
P^{\prime\prime}=c_{\eta z}\frac{2n}{R}.
\end{equation}
Notice that except for the factor of 2, the power $P$ of a mirror Eq.~(\ref{eq:power P prime prime}) is similar to that of a lens in Eq.~(\ref{eq:power P}), which is what we expect.

\section{Conclusions}

In this paper, we used geometric algebra to derive from the unified reflection-refraction law in exponential form the paraxial meridional ray tracing equations in polar form, by equating the arguments of the exponentials and employing the properties of sign functions.  The sign functions, such as the concavity and axial direction functions, make explicit the Cartesian sign convention used, though in symbols and not in words.  We hope that these sign functions would be universally adopted to finally settle the never-ending debate on sign conventions.


\begin{thebibliography}{99}
\footnotesize
\bibitem{NussbaumPhilips_1976_ContemporaryOpticsforScientistsandEngineers_p11-12}
Allen Nussbaum and Richard A. Phillips, \textsl{Contemporary Optics for Scientists and Engineers} (Prentice-Hall, Englewood Cliffs, NJ, 1976), p. 11--12.

\bibitem{Sukheeja_1967_ajp35i7pp611-613}
B. D. Sukheeja, ``Sign conventions in geometrical optics,'' Am. J. Phys. \textbf{35}(7), 611--613 (1967).  

\bibitem{Sukheeja_1969_ajp37i2pp230}
B. D. Sukheeja, ``Clarifications on sign conventions in geometrical optics,'' Am. J. Phys. \textbf{37}(2), 230 (1969).


\bibitem{SandhuFriedman_1969_ajp37i2pp229-230}
H. S. Sandhu and G. B. Friedmann, ''Concerning sign conventions in geometrical optics,'' Am. J. Phys. \textbf{37}(2), 229--230 (1969).

\bibitem{Lemmerhirt_1999_ajp67i5p370}
Fred Lemmerhirt, ``Sign conventions in geometrical optics,'' Am. J. Phys. \textbf{67}(5), 370 (1999).

\bibitem{Nussbaum_2000_ajp68i4p305}
Allen Nussbaum, ``Sign conventions in geometrical optics,'' Am. J. Phys. \textbf{68}(4), 305 (2000).


\bibitem{Welford_1974_AbberationsoftheSymmetricalOpticalSystem_pvi}
W. T. Welford, \textsl{Aberrations of the Symmetrical Optical System} (Academic, London, 1974), p. vi.

\bibitem{Conrady_1985_AppliedOpticsandOpticalDesign}
Alexander Eugen Conrady, \textsl{Applied Optics and Optical Design}, Part I. (Dover, New York, 1985).

\bibitem{SugonMcNamara_2004_ajp72i1pp92-97_p95}
Quirino M. Sugon Jr. and Daniel J. McNamara, ``A geometric algebra reformulation of geometric optics,'' Am. J. Phys. \textbf{72}(1), 92-97.  See pp. 95.

\bibitem{SugonMcNamara_2006_AIEP139pp179-224_p192-195}
Quirino M. Sugon Jr. and Daniel J. McNamara, ''Ray tracing in spherical interfaces using geometric algebra,'' in \textsl{Advances in Imaging and Electron Physics} \textbf{139}, 179--224.  See pp. 192--195.

\bibitem{Hestenes_2003_ajp71i2pp104-121_p110}
David Hestenes, ``Oersted medal lecture 2002: Reforming the mathematical language of physics,'' Am. J. Phys. \textbf{71}(2), 104--121 (2003).  See p. 110.

\bibitem{SugonMcNamara_2004_ajp72i1pp92-97_p93}
See Ref. \cite{SugonMcNamara_2004_ajp72i1pp92-97_p95}, p. 93.

\bibitem{SugonMcNamara_2006_AIEP139pp179-224_p188-196}
See Ref. \cite{SugonMcNamara_2006_AIEP139pp179-224_p192-195}, pp. 188-196.

\bibitem{SugonMcNamara_2006_AIEP139pp179-224_p215-221}
See Ref. \cite{SugonMcNamara_2006_AIEP139pp179-224_p192-195}, pp. 215-221.

\bibitem{Conrady_1985_AppliedOpticsandOpticalDesign_p7}
See Ref. \cite{Conrady_1985_AppliedOpticsandOpticalDesign}, p. 7.

\bibitem{SugonMcNamara_2006_AIEP139pp179-224_p219}
See Ref. \cite{SugonMcNamara_2006_AIEP139pp179-224_p192-195}, pp. 219.

\bibitem{SugonMcNamara_2006_AIEP139pp179-224_p204}
See Ref. \cite{SugonMcNamara_2006_AIEP139pp179-224_p192-195}, pp. 204.

\end{thebibliography}
\end{document}